# Evaluating LeNet Algorithms in Classification Lung Cancer from Iraq-Oncology Teaching Hospital/National Center for Cancer Diseases

Jafar Abdollahi 1, 2

[1]. *Faculty of Electrical and Computer Engineering, Islamic Azad University, Ardabil, Iran*
[2]. *Young Researchers and Elite Club, Ardabil Branch, Islamic Azad University, Ardabil, Iran*
Email: Ja.abdollahi77@gmail.com

*Abstract*— The advancement of computer-aided detection systems had a significant impact on clinical analysis and decision-making on human disease. Lung cancer requires more attention among the numerous diseases being examined because it affects both men and women, increasing the mortality rate. LeNet, a deep learning model, is used in this study to detect lung tumors. The studies were run on a publicly available dataset made up of CT image data (IQ-OTH/NCCD). Convolutional neural networks (CNNs) were employed in the experiment for feature extraction and classification. The proposed system was evaluated on Iraq-Oncology Teaching Hospital/National Center for Cancer Diseases datasets the success percentage was calculated as 99.51%, sensitivity (93%) and specificity (95%), and better results were obtained compared to the existing methods. Development and validation of algorithms such as ours are important initial steps in the development of software suites that could be adopted in routine pathological practices and potentially help reduce the burden on pathologists.

## I. INTRODUCTION

The World Health Organization (WHO) reports that 2.09 million cases of lung cancer were diagnosed globally in 2018. With 1.76 million deaths from this cancer in 2018, it is thought to be the main cause of cancer-related deaths [1]. This is mainly due to tobacco use considered as the most important risk factor for cancer with approximately 22% of cancer deaths [2]. In the USA, researchers estimate about 228,820 new cases and 135,700 deaths in 2020 alone [3].

Segmentation of lungs, (i.e., detecting and isolating the lung lobes), and lung nodules detection (i.e., the detection of presence and the determination of the region of interest of a nodule) is an important procedure in a CAD system when it comes to the diagnosis of malignant lung nodules [4]. Various methods have been proposed including traditional methods [5] as well as deep learning methods [6]. Classical methods for lung segmentation include 3D Region Growing [7], Non-Negative Matrix Factorization with constraints [8], Adaptive Crisp Active Contours [9], Automatic Lung Segmentation using thresholds and morphologic operations [10] etc. Furthermore, lung segmentation methods can be classified in three major categories fowling [11]: deformable boundary-based techniques, edge-based techniques, and threshold-based techniques. For an elaborated comparison of classical techniques, the interested read can refer to [12].

Computer version technology has recently produced some pleasing outcomes. In some research, photographs of renal pathology were processed using computer technology. Most studies used digital slide scanners to convert renal biopsy microscope slides to WSIs initially. Following that, they developed models to handle WSIs, such as object detection on the region of interest (glomeruli, renal tubules, interstitial cells, etc.). On the one hand, the computer version's machine learning algorithms define manually the feature extraction techniques for photos. Medical image and healthcare [13] analysis including whole-slide pathology [14], X-ray [15, 16], diabetes [17, 18], breast cancer [19], heart [20], time series [21], medicinal plants [22], stock market [23], stroke [24], etc. have successfully used machine learning and deep learning.

Kato et al. [25, 26] introduced a method of detecting glomeruli. They used the histogram of oriented gradient (HOG) [27] and segmental-HOG for increasing robustness in the different glomerular morphology to extract features. One of the most common problems and serious illnesses, with the greatest fatality rate, is cancer. Since lung cancer is the most prevalent type of cancer, it accounts for around 25% of all cancer-related deaths. Two major categories can be used to separate the suggested deep learning techniques. The first method entails locating numerous questionable patches that are probably home to a lesion before determining whether a nodule is present utilizing a different module. The second strategy, on the other hand, seeks to systematically identify and categorize nodules. A CAD system based on YOLO (You Only Look Once) [28] is used, for instance, in [29] to identify nodules in CT scans.

3D shape-based descriptors were anticipated by Choi and Choi [30] to identify nodules. Support vector machine (SVM) was used to classify the candidates found using multiscale dot improvement feature descriptors and filtering on the nodules [31]. Shape-based feature descriptors that choose likely candidates using a 3D blob method and accompanying algorithms were proposed by Tas and Ugur [32]. Regarding this, representative characteristics were gathered to distinguish between real nodules and false positives using the SVM classifier. A new method that provides automatic recognition of a juxta-pleural nodule was modelled by Erda Tascet. features were made available to those techniques, which relied on characteristics obtained from texture and form for detection. The significance of investment technique for nodule detection was tested by Santos et al. [33]. (Diameters up to 2–10 mm).

Benign and malignant nodules have considerable feature overlaps, but still must be differentiated based on morphology and location at early stages. Benign nodules are usually located at the peripheral, with smooth surfaces and triangular shapes filled with fat and calcium, while malignant nodules often show speculations with edges, lobulated, vascular convergence, cystic air spaces, pleural indentations, bubble-like lucencies, and sub-solid morphology. Malignancy is also related with size and growth of the nodules [34]. The three different categories (Normal, Malignant, and Bengin) of lung nodules are shown in **Figure 2.** In this study lung patient Computer Tomography (CT) scan images are used to detect and classify the Normal cases, Malignant cases and to detect the Bengin cases. The CT scan images are segmented using LeNet architecture.

**The rest of the article is prepared as follows.** In Section 2, we discuss deep learning methods, and proposed methods. dataset is highlighted in Section 3. The experimental result and discussion are described in Section 4. In Section 5, the conclusion is described.

## II. DEEP LEARNING METHOD

Deep learning, commonly referred to as deep structured learning, is one of several machine learning techniques built on representation learning and artificial neural networks. Unsupervised, semi-supervised, or supervised learning are all possible [35]. In areas like computer vision, speech recognition, natural language processing, machine translation, bioinformatics, drug design, medical image analysis, climate science, material inspection, and board game programs, deep-learning architectures like deep neural networks, deep belief networks, deep reinforcement learning, recurrent neural networks, convolutional neural networks, and Transformers have been applied with results comparable to and I [36, 37, 38].

### A. LeNet Method

In this section, we will introduce LeNet, among the first published CNNs to capture wide attention for its performance on computer vision tasks. The model was introduced by (and named for) Yann LeCun, then a researcher at AT&T Bell Labs, for the purpose of recognizing handwritten digits in images [47]. This work represented the culmination of a decade of research developing technology. In 1989, LeCun's team published the first study to successfully train CNNs via backpropagation [48].

LeNet produced exceptional results at the time, matching the performance of support vector machines, which were at the time the leading supervised learning method, and obtaining an error rate of less than 1% per digit. LeNet was eventually modified to handle deposits in ATMs by recognizing digits. Some ATMs still use the code created in the 1990s by Yann LeCun and his colleague Leon Bottou. LeNet (LeNet-5) is made up of two major components: a dense block with three fully connected layers and a convolutional encoder with two convolutional layers. **Figure 1** depicts a summary of the architecture.

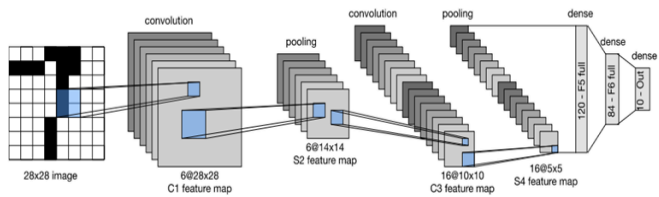

**Fig. 1** Data flow in LeNet. The input is a handwritten digit, the output a probability over 10 possible outcomes.

The basic units in each convolutional block are a convolutional layer, a sigmoid activation function, and a subsequent average pooling operation. Note that while ReLUs and max-pooling work better, these discoveries had not yet been made at the time. Each convolutional layer uses a kernel and a sigmoid activation function. These layers map spatially arranged inputs to several two-dimensional feature maps, typically increasing the number of channels. The first convolutional layer has 6 output channels, while the second has 16. Each pooling operation (stride 2) reduces dimensionality by a factor of via spatial down sampling. The convolutional block emits an output with shape given by (batch size, number of channels, height, width).

To pass output from the convolutional block to the dense block, we must flatten each example in the minibatch. In other words, we take this four-dimensional input and transform it into the two-dimensional input expected by fully connected layers: as a reminder, the two-dimensional representation that we desire uses the first dimension to index examples in the minibatch and the second to give the flat vector representation of each example. LeNet's dense block has three fully connected layers, with 120, 84, and 10 outputs, respectively. Because we are still performing classification, the 10-dimensional output layer corresponds to the number of possible output classes.

## III. IQ-OTH/NCCD - LUNG CANCER DATASET

The lung cancer dataset for the Iraq-Oncology Teaching Hospital/National Center for Cancer Diseases (IQ-OTH/NCCD) was gathered over a three-month period in fall 2019 in the specialty hospitals. It includes CT scans of both healthy volunteers and patients with lung cancer in various stages of the disease. Oncologists and radiologists in these two centers marked IQ-OTH/NCCD slides. 1190 pictures altogether, representing slices from 110 instances' CT scans, are included in the dataset (**see Figure 2**). These cases are divided into three categories: benign, malignant, and normal. Of them, 40 cases have been determined to be malignant, 15 to be benign, and 55 to be normal cases. The first collection of CT images was done in DICOM format.

The scanner is a Siemens SOMATOM model. The CT procedure calls for a slice thickness of 1 mm, a window width of 350 to 1200 HU a, and a window center of 50 to 600 HU a. with a breath-hold throughout inspiration's peak. Before performing analysis, all photos were de-identified. The Oversight Review Board waived the requirement for written consent. The institutional review boards of the participating hospitals gave their approval to the study. Slices are included in every scan. These slices, which range in number from 80 to 200, each depict an image of the human chest with various sides and angles.

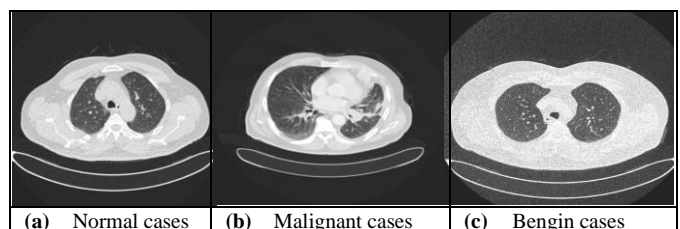

| **(a)** Normal cases | **(b)** Malignant cases | **(c)** Bengin cases |

**Fig 2.** Categories of lung nodules in a CT scan; normal, malignant, and Bengin malignant (from left to right).

The 110 instances differ in terms of gender, age, educational level, locale, and way of life. They include some who work for the oil and transportation ministries in Iraq as well as farmers and gainers. Most of them originate from locations in Iraq's middle, specifically from the provinces of Baghdad, Wasit, Diyala, Salahuddin, and Babylon. Which were split into two non-overlapping sets: a training set (1,097 nodules), used to train the deep learning system and a validation set (197 nodules), used to monitor the performance of the system during training (**Table 1**).

**Table 1** Detailed number of nodules and samples in the training, validation, and test sets in IQ-OTH/NCCD - Lung Cancer Dataset Description.

| Train | | | Test |
|---|---|---|---|
| benign cases | Malignant cases | Normal cases | 197 |
| 120 | 561 | 416 | |

## IV. EXPERIMENTAL RESULTS AND DISCUSSION

### A. Data preprocessing

Pre-processing is one of the key components for classifying histology pictures. While CNNs are often built to receive noticeably smaller inputs, the dataset images are enormous. Therefore, the resolution of the photos must be reduced to capture the input while preserving the essential elements. Data augmentation is used to increase the amount of unique data in the set because the dataset is much smaller than what is typically required to train a DL model correctly. This approach makes a significant contribution to preventing overfitting, a condition where the model correctly absorbs the training data but is entirely unable to classify and generalize unknown images [18, 19, 20].

### B. Training details

We employed pre-existing implementations from well-known classification libraries for this test, where each detection framework's performance was optimized by training hyper-parameter tuning. As a result, we chose to train LeNet using the TensorFlow classification package.

### C. Performance with Different Structure

We conducted controlled experiments and assessed the results using a test dataset. Table 1 displays the model's performance. The harmonic average of recall rates and precision is called F1, or F-measures. The following are the definitions of precision, recall, and F-measures:

**Accuracy**: Indicate the quantity of "correct predictions made" by the category, divided by the quantity of "total predictions made" by a similar category

$$Accuracy = \frac{TP+TN}{TP+FP+TN+FN} \quad (1)$$

**Sensitivity**: Real positive rate: If the result is positive for the person, in a few percent of cases, the model will be positive, which is calculated from the following formula.

$$Sensitivity = \frac{TP}{TP+FN} \quad (2)$$

**Specificity**: If the result is negative for the person, the result will be negative in a few percent of cases. Which is calculated from the following formula [16-26].

$$Specificity = \frac{TN}{TN+FP} \quad (3)$$

Where TP (true positive) represents the number of cases correctly identified as nodules; FP (false positive) represents the number of cases incorrectly identified as nodules; TN (true negative) represents the number of cases correctly identified as non-nodules; and FN (false negative) represents the number of cases incorrectly identified as non-nodules.

### D. Results and Discussion

The deadliest cancer in the world is thought to be lung cancer. Radiologists must examine numerous Computed Tomography (CT) scans to find it. This work is time-consuming and tedious. For the automatic nodule detection and categorization, innovative methods based on deep learning object detection algorithms have recently been developed. These methods can be used to create Computed Aided Detection (CAD) software, which will lighten the workload of radiologists and hasten the screening procedure. **Figure 4** shows our achieved accuracy.

We measure the effectiveness of the focal loss function in lung nodule classification by Lenet. **Figure 3** summarizes our finding in this experiment. We observe that focal loss function can boost the classification's accuracy up to 97.88%, sensitivity up to 93%, and specificity up to 95% in comparison with cross-entropy loss function.

**Table 2.** Performance of The Classifiers

| Performance LeNet Method | |
|---|---|
| Loss | 0.0509 |
| Accuracy | 0.9788 |
| Val Loss | 177.399 |
| Val Acc | 0.8317 |
| Sensitivity | 0.9314 |
| Specificity | 0.9591 |

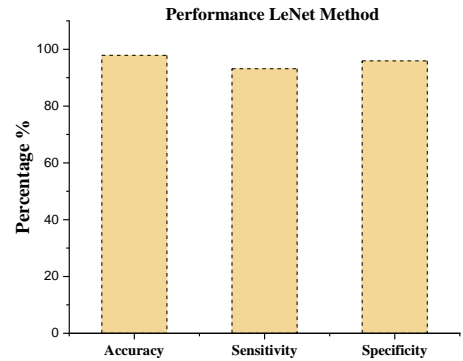

Fig. 3 summarizes our findings in this experiment.

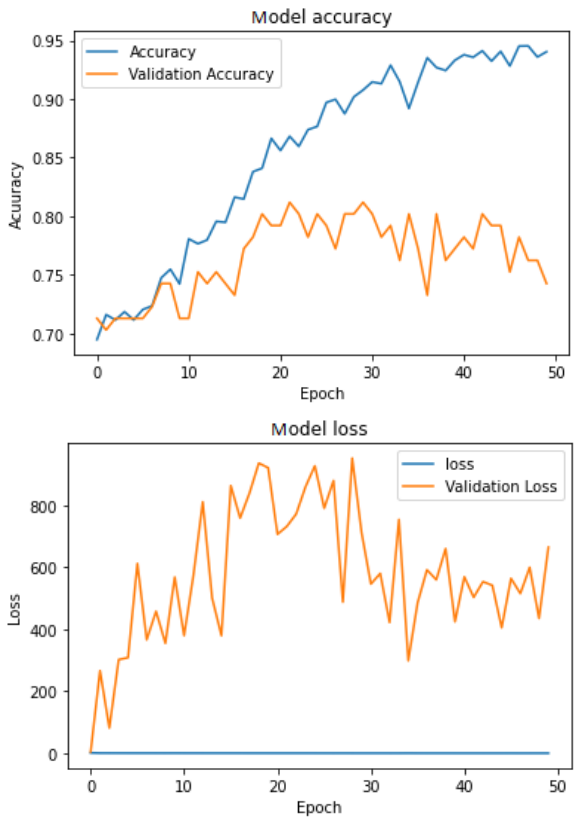

Fig 4. The lines represent the loss (a) and the accuracy (b) of MIL (blue) and CT training method (orange) for the training set (dotted) and the validation set (solid). The x axis represents the numbers of elapsed training epochs.

To evaluate the correctness of our classifier, we use three metrics: accuracy, sensitivity, and specificity. Table 4 presents the comparison of our proposed CNN network with other related works in the literature. From the table, our method has the best sensitivity of 93.14% indicating the high quality of our classifier in determining the pulmonary nodule cases correctly. Besides, although our work has higher accuracy (97.88% vs. 97.6%) and lower specificity (95.91% vs. 100%) than the works in [39, 40], respectively, the consensus level (the number of radiologists agrees with the annotation) used in our work is higher than the ones used in [39, 40]. Our best experimental results (in bold) in **Table 3** prove that our CNN method with focal loss is a high-quality classifier with an accuracy of 97.88%, sensitivity of 93.14%, and specificity of 95.91%.

Table 3. Classification results of LeNet compared with other works.

| Work | Accuracy | Sensitivity | Specificity |
|---|---|---|---|
| Li et al. 2016 [41] | 86.04% | 87.01% | - |
| Choi and Choi 2013 [39] | 97.06% | 95.02% | 96.02% |
| Kuruvilla and Gunavathi 2014 [40] | 93.03% | 91.04% | 100% |
| Shikun Zhang et al, 2019 [42] | 97.04% | 96.68% | |
| Shanchen Pang et al, 2019 [43] | 89.85% | | |
| Iftikhar Naseer et al, 2022 [44] | 97.42% | 97.25% | 97.58% |
| MichailTsivgoulis et al, 2022 [45] | 94.03% | 95.03% | 91.03% |
| This Paper | 97.88% | 93.14% | 95.91% |

## V. CONCLUSION

Numerous cutting-edge deep learning classification techniques have recently been proposed in the literature. At the time of publication, they are primarily weighed against one another using actual photos. These detectors have, however, been effectively used in numerous different fields. However, the task gets more difficult in some applications, such as the categorization of lung nodules, and little effort has been presented to benchmark these methods. It is challenging to compare some detectors because they have already been employed in lung nodule detection solutions with various metrics [46].

In this paper, we presented LeNet algorithms for cancer classification. Results show that LeNet gives the highest accuracy (97.88%) with the lowest error rate. This, in our opinion, can assist in establishing the standards for CAD software development in the future employing deep object identification frameworks. Future research will test the methods using additional deep detectors, additional feature extractors, advanced hyper-parameter optimizations, evaluation of detection performance in three dimensions, and comparison of the methods using different datasets for lung cancer.

### Data Availability

The data used to support the findings of this study are available from the corresponding author upon request (https://data.mendeley.com/datasets/).

### Conflicts of Interest

The authors declare that they have no conflicts of interest to report regarding the publication of this paper.

### REFERENCES

[1] World Health Organization, "Cancer," https://www.who.int/newsroom/fact-sheets/detail/cancer, 9 2018, online; accessed January 2020.
[2] A. Ahmad Kiadaliri, B. Norrving, and GBD 2015 Risk Factors Collaborators, "Global, regional, and national comparative risk assessment of 79 behavioural, environmental and occupational, and metabolic risks or clusters of risks, 1990-2015: a systematic analysis for the global burden of disease study 2015," The Lancet, vol. 388, no. 10053, pp. 1659–1724, 10 2016.
[3] American Cancer Society, "Cancer Facts & Figures 2020," https://www.cancer.org/content/dam/cancer-org/research/cancerfacts-and-statistics/annual-cancer-facts-and figures/2020/cancer-facts and-figures-2020.pdf, American Cancer Society, Tech. Rep., 2020, online; accessed January 2020.
[4] Lee S.L.A., Kouzani A.Z., Hu E.J. Automated detection of lung nodules in computed tomography images: A review Machine Vision and Applications, 23 (1) (2012), pp.
[5] Kamble B., Sahu S.P., Doriya R. A review on lung and nodule segmentation techniques Kolhe M.L., Tiwari S., Trivedi M.C., Mishra K.K. (Eds.), Advances in data and information sciences (vol. 94, 555–565), Springer Singapore (2020)
[6] Gite S., Mishra A., Kotecha K. Enhanced lung image segmentation using deep learning Neural Computing and Applications (2022),
[7] Medeiros da Nobrega, R. V., Rodrigues, M. B., & Filho, P. P. R. (2017). Segmentation and Visualization of the Lungs in Three Dimensions Using 3D Region Growing and Visualization Toolkit in CT Examinations of the Chest. In 2017 IEEE 30th International symposium on computer-based medical systems (pp. 397–402).
[8] Hosseini-Asl E., Zurada J.M., Gimelfarb G., El-Baz A. 3-D lung segmentation by incremental constrained nonnegative matrix factorization IEEE Transactions on Biomedical Engineering, 63 (5) (2016), pp. 952-963,

[9] Rebouças Filho P.P., Cortez P.C., da Silva Barros A.C., Albuquerque V.H.C., Tavares J.M.R.S. Novel and powerful 3D adaptive crisp active contour method applied in the segmentation of CT lung images Medical Image Analysis, 35 (2017), pp. 503-516,

[10] Nery, F., Silva, J. S., Ferreira, N. C., & Caramelo, F. (2012). 3D automatic lung segmentation in low-dose CT. In 2012 IEEE 2nd Portuguese meeting in bioengineering (pp. 1–4).

[11] Shaukat F., Raja G., Frangi A.F. Computer-aided detection of lung nodules: A review Journal of Medical Imaging, 6 (02) (2019), p. 1,

[12] Shaukat F., Raja G., Frangi A.F. Computer-aided detection of lung nodules: A review Journal of Medical Imaging, 6 (02) (2019), p. 1, 10.1117/1.JMI.6.2.020901.

[13] Abdollahi, J., Nouri-Moghaddam, B., & Ghazanfari, M. (2021). Deep Neural Network Based Ensemble learning Algorithms for the healthcare system (diagnosis of chronic diseases). arXiv preprint arXiv:2103.08182. https://doi.org/10.48550/arXiv.2103.08182

[14] Abdollahi, J., Davari, N., Panahi, Y., & Gardaneh, M. (2022). Detection of metastatic breast cancer from whole-slide pathology images using an ensemble deep-learning method. Archives of Breast Cancer.

[15] S. Ramachandran, J. George, S. Skaria, and V. V.V., "Using YOLO based deep learning network for real time detection and localization of lung nodules from low dose CT scans," in Medical Imaging 2018: Computer-Aided Diagnosis, K. Mori and N. Petrick, Eds. Houston, United States: SPIE, Feb. 2018, p. 53.

[16] Abdollahi, J., & Mahmoudi, L. (2022, February). An Artificial Intelligence System for Detecting the Types of the Epidemic from X-rays: Artificial Intelligence System for Detecting the Types of the Epidemic from X-rays. In 2022 27th International Computer Conference, Computer Society of Iran (CSICC) (pp. 1-6). IEEE. DOI: 10.1109/CSICC55295.2022.9780523

[17] Abdollahi, J., Moghaddam, B. N., & Parvar, M. E. (2019). Improving diabetes diagnosis in smart health using genetic-based Ensemble learning algorithm. Approach to IoT Infrastructure. Future Gen Distrib Systems J, 1, 23-30.

[18] Abdollahi, J., Nouri-Moghaddam, B. Hybrid stacked ensemble combined with genetic algorithms for diabetes prediction. Iran J Comput Sci (2022). https://doi.org/10.1007/s42044-022-00100-1

[19] Abdollahi, J., Keshandehghan, A., Gardaneh, M., Panahi, Y., & Gardaneh, M. (2020). Accurate detection of breast cancer metastasis using a hybrid model of artificial intelligence algorithm. Archives of Breast Cancer, 22-28. https://doi.org/10.32768/abc.20207122-28

[20] Abdollahi, J., Nouri-Moghaddam, B. A hybrid method for heart disease diagnosis utilizing feature selection-based ensemble classifier model generation. Iran J Comput Sci (2022). https://doi.org/10.1007/s42044-022-00104-x

[21] Abdollahi, J., Irani, A. J., & Nouri-Moghaddam, B. (2021). Modeling and forecasting Spread of COVID-19 epidemic in Iran until Sep 22, 2021, based on deep learning. arXiv preprint arXiv:2103.08178. https://doi.org/10.48550/arXiv.2103.08178

[22] Abdollahi, J. (2022, February). Identification of Medicinal Plants in Ardabil Using Deep learning: Identification of Medicinal Plants using Deep learning. In 2022 27th International Computer Conference, Computer Society of Iran (CSICC) (pp. 1-6). IEEE. DOI: 10.1109/CSICC55295.2022.9780493

[23] Abdollahi, J., & Mahmoudi, L. Investigation of artificial intelligence in stock market prediction studies. 10 th International Conference on Innovation and Research in Engineering Science

[24] Amani F, Abdollahi J, mohammadnia alireza, amani paniz, fattahzadeh-ardalani ghasem. Using Stacking methods based Genetic Algorithm to predict the time between symptom onset and hospital arrival in stroke patients and its related factors. JBE. 2022;8(1):8-23.

[25] Kakimoto, T.; Okada, K.; Hirohashi, Y.; Relator, R.; Kawai, M.; Iguchi, T.; Fujitaka, K.; Nishio, M.; Kato, T.; Fukunari, A.; et al. Automated image analysis of a glomerular injury marker desmin in spontaneously diabetic Torii rats treated with losartan. J. Endocrinol. 2014, 222, 43-51.

[26] Kato, T.; Relator, R.; Ngouv, H.; Hirohashi, Y.; Takaki, O.; Kakimoto, T.; Okada, K. Segmental HOG: New descriptor for glomerulus detection in kidney microscopy image. BMC Bioinform. 2015, 16, 316.

[27] B. van Ginneken, A. A. A. Setio, C. Jacobs, and F. Ciompi, "Off-the-shelf convolutional neural network features for pulmonary nodule detection in computed tomography scans," in 2015 IEEE 12th International Symposium on Biomedical Imaging (ISBI), Apr. 2015, pp. 286–289, iSSN: 1945-8452.

[28] J. Redmon, S. Divvala, R. Girshick, and A. Farhadi, "You Only Look Once: Unified, Real-Time Object Detection," in 2016 IEEE Conference on Computer Vision and Pattern Recognition (CVPR). Las Vegas, NV, USA: IEEE, Jun. 2016, pp. 779–788.

[29] S. Ramachandran, J. George, S. Skaria, and V. V.V., "Using YOLO based deep learning network for real time detection and localization of lung nodules from low dose CT scans," in Medical Imaging 2018: Computer-Aided Diagnosis, K. Mori and N. Petrick, Eds. Houston, United States: SPIE, Feb. 2018, p. 53.

[30] Choi, W. J., & Choi, T. S. (2014). Automated pulmonary nodule detection based on three-dimensional shape-based feature descriptor. Computer methods and programs in biomedicine, 113(1), 37-54.

[31] Campadelli, P., Casiraghi, E., & Valentini, G. (2005). Support vector machines for candidate nodules classification. Neurocomputing, 68, 281-288.

[32] Taşçı E, Ugur A. Shape and texture based novel features for automated Juxtapleural nodule detection in lung CTs. J Med Syst. 2015;39(5):1-13.

[33] Santos AM, de Carvalho Filho AO, Silva AC, et al. Automatic detection of small lung nodules in 3D CT data using Gaussian mixture models, Tsallis entropy and SVM. Eng Appl Artif Intell. 2014;36:27-39.

[34] Peña DM, Luo S, Abdelgader A. Auto diagnostics of lung nodules using minimal characteristics extraction technique. Diagnostics. 2016; 6:13.

[35] LeCun, Yann; Bengio, Yoshua; Hinton, Geoffrey (2015). "Deep Learning". Nature. 521 (7553): 436–444. Bibcode:2015Natur.521..436L. doi:10.1038/nature14539. PMID 26017442. S2CID 3074096.

[36] Ciresan, D.; Meier, U.; Schmidhuber, J. (2012). "Multi-column deep neural networks for image classification". 2012 IEEE Conference on Computer Vision and Pattern Recognition. pp. 3642–3649. arXiv:1202.2745

[37] Krizhevsky, Alex; Sutskever, Ilya; Hinton, Geoffrey (2012). "ImageNet Classification with Deep Convolutional Neural Networks" (PDF). NIPS 2012: Neural Information Processing Systems, Lake Tahoe, Nevada. Archived (PDF) from the original on 2017-01-10. Retrieved 2017-05-24.

[38] Google's AlphaGo AI wins three-match series against the world's best Go player". TechCrunch. 25 May 2017. Archived from the original on 17 June 2018. Retrieved 17 June 2018.

[39] W.-J. Choi and T.-S. Choi, "Automated pulmonary nodule detection system in computed tomography images: a hierarchical block classification approach," Entropy, vol. 15, no. 2, pp. 507–523, 2013.

[40] J. Kuruvilla and K. Gunavathi, "Lung cancer classification using neural networks for CT images," Computer Methods and Programs in Biomedicine, vol. 113, no. 1, pp. 202–209, 2014.

[41] W. Li, P. Cao, D. Zhao, and J. Wang, "Pulmonary nodule classification with deep convolutional neural networks on computed tomography images," Computational and Mathematical Methods in Medicine, vol. 2016, Article ID 6215085, 7 pages, 2016.

[42] Zhang, S., Sun, F., Wang, N., Zhang, C., Yu, Q., Zhang, M., ... & Zhong, H. (2019). Computer-aided diagnosis (CAD) of pulmonary nodule of thoracic CT image using transfer learning. Journal of digital imaging, 32(6), 995-1007.

[43] Pang, S., Zhang, Y., Ding, M., Wang, X., & Xie, X. (2019). A deep model for lung cancer type identification by densely connected convolutional networks and adaptive boosting. IEEE Access, 8, 4799-4805.

[44] Naseer, I., Akram, S., Masood, T., Jaffar, A., Khan, M. A., & Mosavi, A. (2022). Performance analysis of state-of-the-art CNN architectures for luna16. Sensors, 22(12), 4426.

[45] Tsivgoulis, M., Papastergiou, T., & Megalooikonomou, V. (2022). An improved SqueezeNet model for the diagnosis of lung cancer in CT scans. Machine Learning with Applications, 10, 100399.

[46] D. Riquelme and M. A. Akhloufi, "Deep Learning for Lung Cancer Nodules Detection and Classification in CT Scans," AI, vol. 1, no. 1, pp. 28–67, 2020.

[47] LeCun, Y., Bottou, L., Bengio, Y., Haffner, P., & others. (1998). Gradient-based learning applied to document recognition. Proceedings of the IEEE, 86(11), 2278–2324.

[48] LeCun, Y., Boser, B., Denker, J. S., Henderson, D., Howard, R. E., Hubbard, W., & Jackel, L. D. (1989). Backpropagation applied to handwritten zip code recognition. Neural computation, 1(4), 541–551